# Reliable Robust and Real-Time Communication Protocol for Data Delivery in Wireless sensor Networks

Deepali Virmani [1] , Satbir Jain [2]

[1] Department of computer science, BPIT, IPU, Delhi, India

[2] Department of computer science, NSIT, DU, Delhi, India

deepalivirmani@gmail.com

## Abstract

WSNs can be considered a distributed control system designed to react to sensor information with an effective and timely action. For this reason, in WSNs it is important to provide real-time coordination and communication to guarantee timely execution of the right actions. In this paper a new communication protocol RRRT to support robust real-time and reliable event data delivery with minimum energy consumption and with congestion avoidance in WSNs is proposed. The proposed protocol uses the fault tolerant optimal path for data delivery. The proposed solution dynamically adjust their protocol configurations to adapt to the heterogeneous characteristics of WSNs. Specifically, the interactions between contention resolution and congestion control mechanisms as well as the physical layer effects in WSNs are investigated.

## 1. Introduction

The existing and potential applications of WSNs span a very wide range, including real time target tracking and surveillance, homeland security, and biological or chemical attack detection [1]. Realization of these currently designed and envisioned applications, however, directly depends on real-time and reliable communication capabilities of the deployed sensor/sub-sink network. In this paper, a reliable, robust and real-time communication RRRT protocol is proposed to address the communication challenges introduced by the coexistence of sensors and sub-sinks in WSNs. The RRRT protocol is a novel transport solution that seeks to achieve reliable and timely event detection with minimum possible energy consumption. The RRRT uses a fault tolerant optimal path (FTOP) [2] for data delivery. It includes a combined congestion control mechanism that serves the dual purpose of achieving reliability and conserving energy. The RRRT protocol operation is determined by the current network state based on the delay constrained event reliability and congestion condition in the network. If the delay constrained event reliability is lower than required, RRRT adjusts the reporting frequency of source nodes aggressively to reach the desired reliability level as soon as possible. If the reliability is higher than required, then RRRT reduces the reporting frequency conservatively to conserve energy while still maintaining reliability. This self configuring nature of RRRT makes it robust to random, dynamic topology in WSNs. Furthermore, to address the different reliability requirements of sub-sink and sub-sink communication, RRRT incorporates adaptive rate-based transmission control and (SACK)-based reliability mechanism during sub-sink and sub-sink communication. Performance evaluation via simulation experiments shows that RRRT achieves high performance in terms of reliable event detection, communication latency and energy consumption in WSNs.

## 2. RRRT Protocol Design Principles

Unlike traditional networks, the sensor/sub-sink network paradigm necessitates that the event features are collaboratively estimated within a certain reliability and real-time delay bound. To achieve this objective with maximum resource efficiency, the RRRT protocol exploits both the correlation and the collaborative nature of the network. In the following sections, we first describe the characteristics and challenges of both sensor/sub-sink and sub-sink/sub-sink communication and then based on these characteristics, we discuss the main design components of the RRRT protocol in detail.

### 2.1 Reliable Event Transport

The RRRT protocol is equipped with different reliability functionalities to address heterogeneous requirements of both sensor/sub-sink and sub-sink/sub-sink communication. Next, the main features of these reliability functionalities are described.

**2.1.1 Sensor/Sub-sink Transport Reliability**

In WSNs, sensor/sub-sink transport is characterized by the dense deployment of sensors that continuously observe physical phenomenon. Because of the high density in the network topology, sensor observations are highly correlated in the space domain. In addition, the nature of the physical phenomenon constitutes the temporal correlation between each consecutive observation of the sensor. Because of these spatial and temporal correlations along with the collaborative nature of the WSNs, sensor/sub-sink transport does not require 100% reliability [3], [4].

The RRRT protocol also considers the new notion of event-to-action delay bound to meet the application-specific deadlines. Based on both event transport reliability and event-to-action delay bound notions, we introduce the following definitions:

1. The observed delay-constrained event reliability ($DR_o$) is the number of received data packets within a certain delay bound at the sub-sink node in a decision interval i. In other words, $DR_o$ counts the number of correctly received packets complying with the application-specific delay bounds and the value of $DR_o$ is measured in each decision interval i.

2. The desired delay-constrained event reliability ($DR_d$) is the minimum number of data packets required for reliable event detection within a certain application specific delay bound. This lower bound for the reliability level is determined by the application and based on the physical characteristics of the event signal being tracked.

3. The delay-constrained reliability indicator ($\alpha$) is the ratio of the observed and desired delay-constrained event reliabilities, i.e $\alpha = \dfrac{DR_o}{DR_d}$ \quad (1)

Based on the packets generated by the sensor nodes in the event area, the event features are estimated and $DR_o$ is observed at each decision interval i to determine the necessary action. If the observed delay constrained event reliability is higher than the reliability bound, i.e., $DR_o > DR_d$, then the event is deemed to be reliably detected within a certain delay bound. Otherwise, appropriate action needs to be taken to assure the desired reliability level in sensor/sub-sink communication.

### 2.1.2 Sub-sink-Sub-sink Transport Reliability

In WSNs, a reliable and timely sub-sink-sub-sink ad hoc communication is also required to collaboratively perform the right action upon the sensed phenomena [1]. The RRRT protocol simultaneously incorporates adaptive rate-based transmission control and (SACK)-based reliability mechanism to achieve 100% packet reliability in the required ad hoc communication. To achieve this objective, RRRT protocol relies upon new feedback based congestion control mechanisms and probe packets to recover from subsequent losses and selective-acknowledgments (SACK) to detect any holes in the received data stream. These algorithms are shown to be beneficial and effective in recovering from multiple packet losses in one round-trip time (RTT) especially [5].

## 2.2 Real-Time Event Transport

To assure accurate and timely action on the sensed phenomena, it is imperative that the event is sensed, transported to the sub-sink node and the required action is performed within a certain delay bound. We call this event-to-action delay, $\delta_{e2a}$, which is specific to application requirements and must be met so that the overall objective of the sensor/sub-sink sub-sink network is achieved. The event-to-action delay $\delta_{e2a}$, has three main components as outlined below:

1. Event transport delay $ET_{del}$: It is mainly defined as the time between when the event occurs and when it is reliably transported to the sub-sink node. In general, it involves the following delay components:

(a) Buffering delay ($B_{del}$): It is the time spent by a data packet in the routing queue of an intermediate forwarding sensor node i. It depends on the current network load and transmission rate of each sensor node.

(b) Channel access delay ($CA_{del}$): It is the time spent by the sensor node i to capture the channel for transmission of the data packet generated by the detection of the event. It depends on the channel access scheme in use, node density and the current network load.

(c) Transmission delay ($T_{del}$): It is the time spent by the sensor node i to transmit the data packet over the wireless channel. It can be calculated using transmission rate and the length of the data packet.

(d) Propagation delay ($P_{del}$): It is the propagation latency of the data packet to reach the next hop over the wireless channel. It mainly depends on the distance and channel conditions between the sender and receiver.

2. Event processing delay ($EP_{del}$): This is the processing delay experienced at the sub-sink node when the desired features of event are estimated using the data packets received from the sensor field. This may include a certain decision interval [3] during which the sub-sink node waits to receive adequate samples from the sensor nodes.

3. Action delay ($A_{del}$): The action delay is the time it takes from the instant that event is reliably detected at the sub-sink node to the instant that the actual action is taken. It is composed of the task assignment delay, i.e., time to select the best set of sub-sinks for the task and the action execution delay, i.e., time to actually perform the action. For a timely action it is necessary that the following relation holds:

$$\delta_{e2a} \geq B_{del} + EP_{del} + A_{del} \qquad (2)$$

## 2.3 Congestion Detection and Control Mechanism

In WSNs, because of the memory limitations of the sensor nodes and limited capacity of shared wireless medium, congestion might be experienced in the network. Congestion leads to both waste of communication and energy resources of the sensor nodes and also hampers the event detection reliability because of packet losses [3]. Hence, it is mandatory to address the congestion in the sensor field to achieve real-time and reliable event detection and minimize energy consumption. Only the sub-sink node, and not any of the sensor nodes, can determine the delay-constrained reliability indicator, $\alpha = \dfrac{DR_o}{DR_d}$ and act accordingly.

## 3. RRRT Protocol Operation for Sensor-Sub-sink Communication

In this section, we describe the RRRT protocol operation during sensor/sub-sink communication. Recall that in the previous sections, based on the delay-constrained event reliability and the event-to-action delay bound notions, we had defined a new delay-constrained reliability indicator $\alpha = \dfrac{DR_o}{DR_d}$, i.e., the ratio of observed and desired delay-constrained event reliabilities. To determine proper event reporting frequency update policies, we also define $T_i$ and $T_{sa}$, which are the amount of time needed to provide delay-constrained event reliability for a decision interval i and the application specific sensor/sub-sink communication delay bound, respectively. In conjunction with the congestion notification information (CN bit) and the values of $f_i$, $\alpha_i$, $T_i$ and $T_{sa}$, the sub-sink node calculates the updated reporting frequency, $f_{i+1}$, to be broadcast to source nodes in each decision interval. This updating process is repeated until the optimal operating point is found, i.e., adequate reliability and no congestion condition are obtained. In the following sections, we describe the details of the reporting frequency update policies and possible network conditions experienced by the sensor nodes.

### 3.1 Early Reliability and No Congestion Condition

In this condition, the required reliability level specific to application is reached before the sensor/sub-sink communication delay bound, i.e., $T_i < T_{sa}$, and no congestion is observed in the network, i.e., CN = 0. However, the observed delay-constrained event reliability, $DR_o$, is larger than desired delay-constrained event reliability, $DR_d$ This is because source nodes transmit event data more frequently than required. The most important consequence of this condition is excessive energy consumption of the sensors. Therefore, the reporting frequency should be decreased cautiously to conserve energy. This reduction should be performed cautiously so that the delay-constrained event reliability is always maintained. Therefore, the sub-sink node decreases the reporting frequency in a controlled manner. Intuitively, we try to find a balance between saving energy and maintaining reliability. Hence, the updated reporting frequency can be expressed as follows:

$$f_{i+1} = f_i \frac{T_i}{T_{sa}} \quad (3)$$

## 3.2 Early Reliability and Congestion Condition

In this condition, the required reliability level specific to application is reached before the sensor/sub-sink communication delay bound, i.e., $T_i < T_{sa}$, and congestion is observed in the network, i.e., CN = 1. However, the observed delay-constrained event reliability, $DR_o$, is larger than the desired delay-constrained event reliability, $DR_d$. In this situation, the RRRT protocol decreases reporting frequency to avoid congestion and save the limited energy of sensors. This reduction should be in a controlled manner so that the delay-constrained event reliability is always maintained. However, the reporting frequency can be decreased more aggressively than the case where there is no congestion and the observed delay-constrained event reliability, $DR_o$, is larger than the desired delay-constrained event reliability, $DR_d$. This is because in this case, we are farther from optimal operating point. Here, we try to avoid congestion as soon as possible. Hence, the updated reporting frequency can be expressed as follows:

$$f_{i+1} = \min\left( f_i \frac{T_i}{T_{sa}}, f_i^{\frac{T_i}{T_{sa}}} \right) \qquad (4)$$

## 3.3 Low Reliability and No Congestion Condition

In this condition, the required reliability level specific to application is not reached before sensor-sub-sink communication delay bound, i.e. $T_i > T_{sa}$, and no congestion is observed in the network, i.e., CN = 0. However, the observed delay-constrained event reliability $DR_o$, is lower than the desired delay-constrained event reliability, $DR_d$. The RRRT protocol can work with any of these routing schemes. Therefore, to achieve required event reliability, we need to increase the data reporting frequencies of source nodes. Here, we exploit the fact that the $Dr$ vs $f$ relationship in the absence of congestion, i.e., for $f < f_{max}$ is linear. In this regard, we use the multiplicative increase strategy to calculate updated reporting frequency, which is expressed as follows:

$$f_{i+1} = f_i \frac{DR_d}{DR_o} \qquad (5)$$

## 3.4 Low Reliability and Congestion Condition

In this condition, the required reliability level specific to application is not reached before sensor-sub-sink communication delay bound, i.e., $T_i > T_{sa}$, and congestion is observed in the network, i.e., CN = 1. However, the observed delay-constrained event reliability, $DR_o$, is lower than the desired delay-constrained event reliability, $DR_d$. This situation is the worst possible case, since desired delay-constrained event reliability is not reached, network congestion is observed and thus, limited energy of sensors is wasted. Hence, the RRRT protocol aggressively reduces reporting frequency to reach optimal reporting frequency as soon as possible. Therefore, to assure sufficient decrease in the reporting frequency, it is exponentially decreased and the new frequency is expressed by:

$$f_{i+1} = f_i^{\frac{DR_o}{(DR_d * X)}} \qquad (6)$$

Where x denotes the number of successive decision intervals for which the network has remained in the same situation including the current decision interval, i.e. $x \geq 1$. Here, the purpose is to decrease reporting frequency with greater aggression, if a network condition transition is not detected.

## 3.5 Adequate Reliability and No Congestion Condition

In this condition, the network is within β tolerance of the optimal operating point, i.e., $f < f_{max}$ and, $1 - \beta \leq \delta_i \leq 1 + \beta_i$ and no congestion is observed in the network. Hence, the reporting frequency of source nodes is left constant for the next decision interval:

$$f_{i+1} = f_i \qquad (7)$$

Here, our aim is to operate as close to $\delta_i = 1$ as possible, while utilizing minimum network resources and meeting event delay bounds. For practical purposes, we define a tolerance level, β, for optimal operating point. The entire RRRT protocol operation is presented in the pseudo-algorithm given in Figure 1

```
        x=1 ;
     RRRT()
       If (Congestion)
        If ( δ < 1 )
       /* Low Reliability and Congestion */
```
$$f_{i+1} = f_i^{\frac{DR_o}{(DR_d * X)}} ;$$
$$x = x+1 ;$$
```
      else if  (δ > 1)
        /* Early Reliability and Congestion */
            x = 1 ;
```
$$f_{i+1} = \min\left(f_i \frac{T_i}{T_{sa}}, f_i^{\frac{T_i}{T_{sa}}}\right);$$
```
          end;
       else if  (No Congestion)
              x = 1;
          if  (δ < 1− β )
   /* Low Reliability and No Congestion */
```
$$f_{i+1} = f_i \frac{DR_d}{DR_o} ;$$
```
     else if ( δ > 1+ β )
    /* Early Reliability and No Congestion */
```
$$f_{i+1} = f_i \frac{T_i}{T_{sa}}; \quad \text{end};$$
```
   else if  1− β ≤ δ_i ≤ 1+ β_i
   /* Adequate Reliability and No Congestion */
```
$$f_{i+1} = f_i ;$$
```
end;
end;
```

Figure 1 Algorithm of RRRT

## 4. RRRT Protocol Operation for Sub-sink/Sub-sink Communication

In this section, we describe the protocol operation of RRRT during Sub-sink/Sub-sink communication. The protocol operation is composed of two main states: i) Start-up state, ii) Steady state. In Figure 2, the RRRT protocol state diagram for Sub-sink/Sub-sink communication is shown.

The operations at each state are described in detail.

**1. Start-Up State:** When establishing new connection between sender and receiver, the sender transports a probe packet towards the receiver to capture the available transmission rate quickly. Each intermediate node between the sender and receiver intercepts the probe packet and updates the bottleneck delay field of the probe packet, if the current value of delay information is higher than that of the intermediate node.

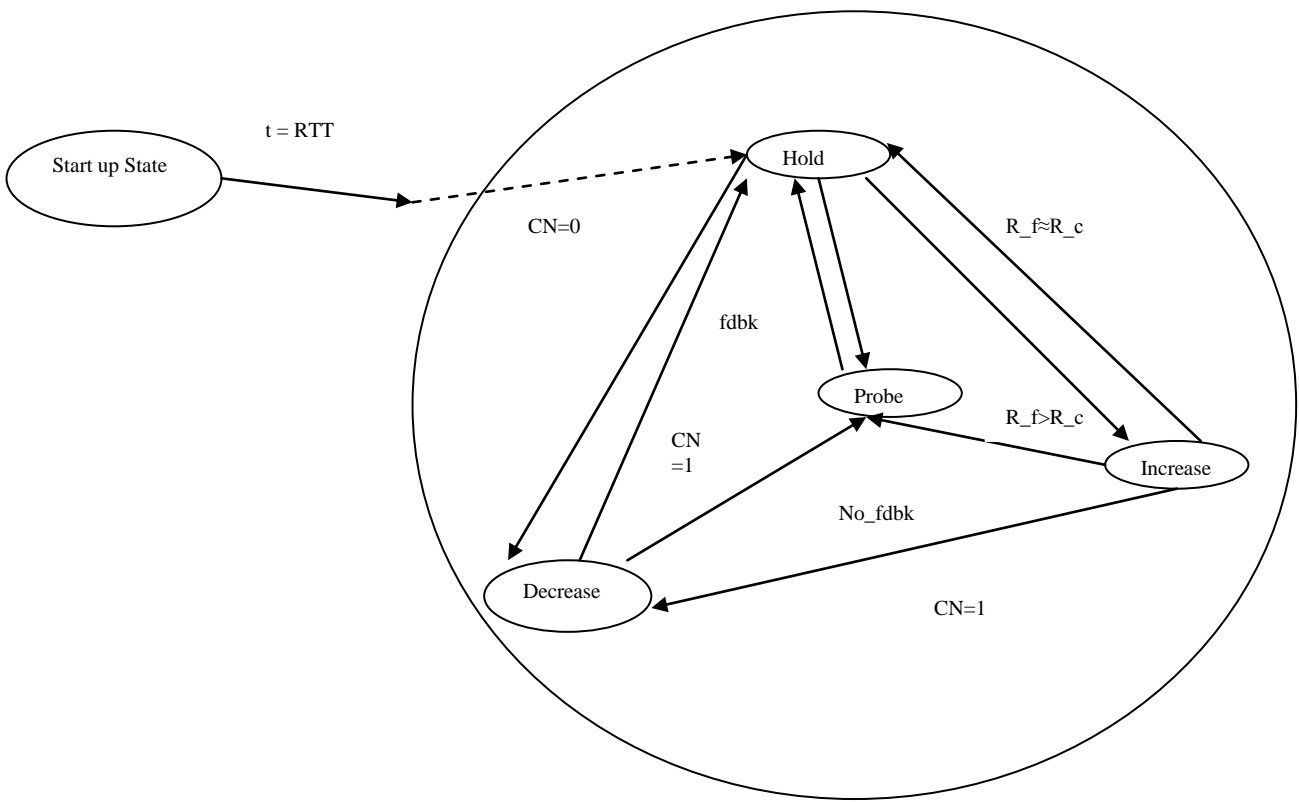

Figure2 RRRT protocol state diagram for Sub-sink/Sub-sink communication

Initially, the delay value of probe packet is assigned to zero. Therefore, after one round-trip-time, the sender gets estimated rate feedback from the receiver, which results in quick convergence to available transmission rate. Furthermore, this probing mechanism of start up phase is also applied after route changes.

**2. Steady State**: This state consists of four sub states: i) Increase, ii) Decrease, iii) Hold and iv) Probe. In the following, we describe the RRRT protocol operations in each sub state:

(a) **Increase:** In this state, the sender increases its transmission rate according to the feedback coming from the receiver. Once an increase decision for sender transmission rate is taken, only $m$ fraction of the difference between transmission rate feedback (R_f) and sender current transmission rate (R_c) is performed. The appropriate fraction value ($m$) for the transmission rate increase is obtained as follows: If the hop count along the data path is greater than or equal to 4 for that connection, m is set to 4. Otherwise, if the hop count is less than 4, then m is set to the actual hop count value along the path. The inherent spatial reuse property of underlying CSMA/CA based MAC protocol requires this normalization in transmission rate.

(b) **Decrease:** In this state, the sender reduces its transmission rate according to the feedback coming from the receiver. Note that the transmission rate is decreased until the minimum transmission rate ($R_{min}$) is reached. $R_{min}$ represents the minimum transmission rate requirement to transfer a certain amount of data within event-to-action delay bound. $R_{min}$ can be calculated as follows:

$$R_{min} = \frac{B}{\Delta_{re2a}}$$

where B represents the amount of packets that should be transmitted to the sub-sink and $\Delta_{re2a}$ is remaining event-to-action deadline.

(c) **Hold:** In this state, the required transmission rate is reached. Sender does not change the transmission rate unless route failure or congestion occurs in the network.

(d) **Probe:** In this state, the sender sends a probe packet to the receiver so as to monitor the available transmission rate in the network as in start up phase.

Overall, the RRRT protocol dynamically shapes data traffic based on both delay bounds and the current conditions of the network. Note that, in the protocol operation, the sender adjusts its transmission rate in response to the rate feedbacks from the receiver, which are sent with the period of $T_{fdbk}$. To prevent the sender from over flooding the network in case all the feedback packets from the receiver are lost, the RRRT protocol also performs a multiplicative decrease of transmission rate for each feedback periods, in which the sender does not receive feedback from the receiver up to a maximum of two feedback periods. After the second feedback period, if the sender still does not receive any feedback packet, it enters into probe state so as to monitor the available transmission rate in the network. In this respect, the periods of feedback $T_{fdbk}$ and probe packets $T_p$ should be larger than one round-trip-time (RTT) and small enough to capture the network dynamics.

## 5. RRRT Performance Evaluation

### 5.1 Sensor/Sub-sink Communication

To evaluate the performance of the RRRT protocol during sensor-sub-sink communication, we developed an evaluation environment using J-Sim [6]. For sensor/sub-sink communication scenario, the number of sources, sensor/sub-sink delay bound and tolerance level were selected as $n = 81$, 1s and $\in = 5\%$, respectively. The event radius was fixed at 45m. We run 10 experiments for each simulation configuration. Each data point on the graphs is averaged over 10 simulation runs. Moreover, in this simulation scenario, the sub-sink nodes, which receive data packets from sensors, stop their movements once they start to receive data.

To further investigate RRRT protocol convergence results, we have compared RRRT protocol, ESRT [3], ATP [5], SPEED [7] protocols in terms of convergence time to (Adequate reliability, No congestion) condition and total energy consumption. The reason for comparison with ESRT and ATP is that both of them are based on event transport reliability notion unlike the other transport layer protocols addressing conventional end-to-end reliability in WSNs. SPEED are a well known real time communication protocol.

As shown in figure 3 and figure 4, the convergence time and total energy consumption of the RRRT protocol are much smaller than those of ESRT, ATP and SPEED for different initial network conditions. This is because ESRT and ATP do not consider application-specific delay bounds while avoiding network congestion and adjusting reporting rate of sensor nodes.

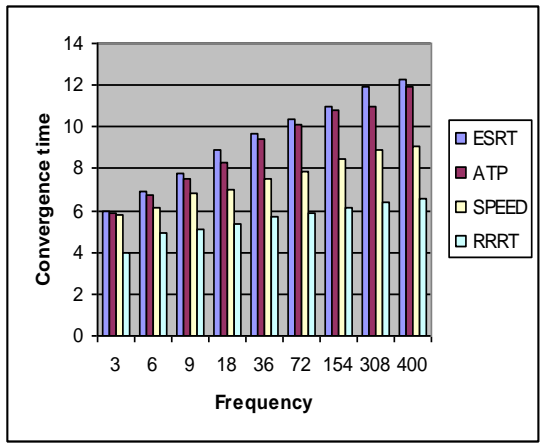
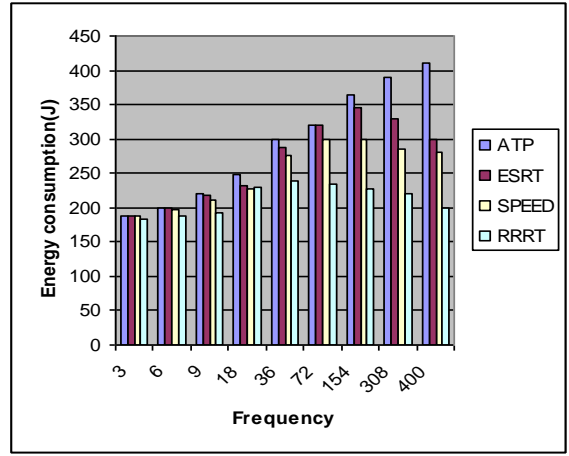

Figure 3 Convergence time          Figure 4 Energy Consumption

## 5.2 Sub-sink/Sub-sink Communication

For sub-sink/sub-sink communication scenario, the performance of the RRRT protocol is evaluated and compared against ESRT [3], ATP [5] and SPEED [7]. The main performance metrics that we employ to measure the performance of the RRRT protocol are aggregate throughput and average packet delay. Here, the aggregate throughput reflects the number of packets successfully received at the destination. By average packet delay, we refer to average latency of data packets during sub-sink/sub-sink communication. All the simulations last for 1000 s. We run 10 experiments for each simulation configuration and each data point on the graphs is averaged over 10 simulation runs.

In Figure 5, we present the aggregate throughput results of the RRRT protocol and other ad hoc transport protocols, i.e. ATP, ESRT and SPEED. In terms of aggregate throughput, the RRRT protocol outperforms other transport protocols under comparison, since RRRT dynamically shapes data traffic according to the channel condition and intermediate node feedbacks. In Figure 6, we also show the average packet delay results of the RRRT and the other transport protocols. As shown in Figure 6, for all simulation configurations, the average packet delay values of RRRT are much lower than those of other protocols, since RRRT captures the available bandwidth in the network quickly and does not allow a burst of packet transmissions with explicit congestion notification and rate feedback based mechanisms

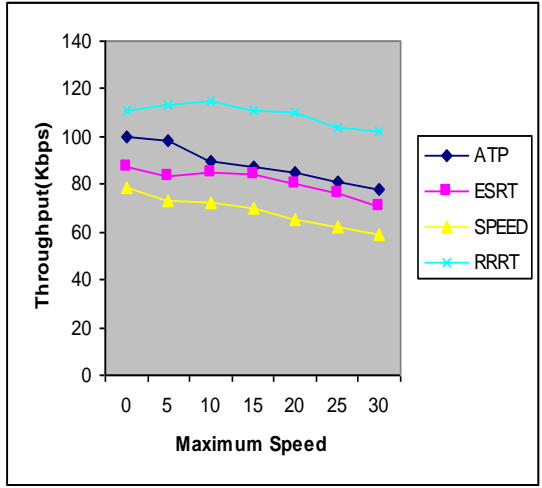 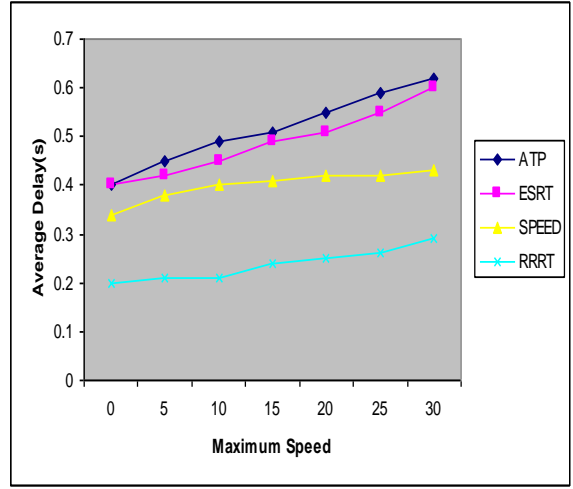

Figure 5 Throughput                Figure 6 Average Delay

## 6. Conclusions

In this paper a real-time and reliable transport RRRT protocol was proposed to address the communication challenges introduced by the coexistence of sensors/sub-sinks in WSNs. The RRRT protocol is a novel transport solution that seeks to achieve reliable and timely event detection with minimum possible energy consumption. It includes a combined congestion control mechanism that serves the dual purpose of achieving reliability and conserving energy. The RRRT protocol operation is determined by the current network state based on the delay-constrained event reliability and congestion condition in the network. The RRRT uses a fault tolerant optimal path for data delivery. If the delay-constrained event reliability is lower than required, RRRT adjusts the reporting frequency of source nodes aggressively to reach the desired reliability level as soon as possible. If the reliability is higher than required, then RRRT reduces the reporting frequency conservatively to conserve energy while still maintaining reliability. This self configuring nature of RRRT makes it robust to random, dynamic topology in WSANs. Furthermore, to address the different reliability requirements of sub-sink-sub-sink communication, RRRT incorporates adaptive rate-based transmission control and (SACK)-based reliability mechanism during sub-sink-sub-sink communication. Performance evaluation via simulation experiments shows that RRRT achieves high performance in terms of reliable event detection, communication latency and energy consumption in WSNs.